\documentstyle[12pt]{article}

\renewcommand{\thefootnote}{\fnsymbol{footnote}}
\newcommand{\putaddress}{\addtocounter{footnote}{+1}
    \footnotetext{Department of Mathematics, SPb UEF,
    Griboyedova 30--32, 191023, St-Petersburg,
    Russia (address for correspondence);
    e-mail: zzz@russ-museum.spb.su}
    \addtocounter{footnote}{+1}
    \footnotetext{Division of Mathematics,
    Istituto per la Ricerca di Base,
    I-86075, Monteroduni (IS), Molise, Italy}}

\newcommand{\algebra}{{\cal A}}
\newcommand{\algspan}{{\rm span}}
\newcommand{\atlas}{{\cal F}}
\newcommand{\cfield}{{\bf C}}
\newcommand{\der}{{\rm Der}}

\newcommand{\ideal}{{\cal I}}
\newcommand{\kernel}{{\rm ker}}
\newcommand{\mata}{{\rm Mat}}
\newcommand{\spat}{{\rm spat}}
\newcommand{\tr}{{\rm tr}}
\newcommand{\vctr}{{\bf v}}
\newcommand{\vertex}{{\circle*{1}}}

\newcommand{\llll}[5]{\put(#1,#2){\line(#3,#4){#5}}}
\newcommand{\mmmm}[8]{\multiput(#1,#2)(#3,#4){#5}{\line(
#6,#7){#8}}}

\date{}
\title{Finitary Algebraic Superspace}
\author{
R.R.Zapatrin\footnotemark[1]{\makebox[0.4em]{}}\footnotemark[2]}

\begin{document}
\unitlength=1mm
\raggedbottom

\maketitle
\putaddress

\begin{abstract}
An algebraic scheme is suggested in which discretized spacetime
turns out to be a quantum observable. As an example, a toy model
producing spacetimes of four points with different topologies is
presented. The possibility of incorporating this scheme into the
framework of non-commutative differential geometry is discussed.
\end{abstract}

\section*{Introduction}

Most of schemes dealing with quantization of gravity are built in
such a way that the geometrical features of the spacetime such as
metric or connection are subject to variation. Rather, on the set
theoretical and topological level the structure of the underlying
spacetime remains unchanged.

It should be emphasized that there is no experimental evidence to
suppose
the set of all spacetime points to be given once and forever.
Nobody can directly observe the points of spacetime and,
therefore, one should not be surprised  when it happens that the
entire topology of the spacetime seems different for different
observers.

The concept of superspace (Misner, Thorne and Wheeler, 1973) was
initially introduced as a space whose points are spacetime
geometries, so their idea could be characterized as a desire to
build a classical model for the kinematics and the dynamics of
spacetime geometry.  In the present paper a quantum analog of the
superspace in which the spacetime itself becomes an observable is
suggested.  To bring this idea to a technical footing I confine
myself by finitary spacetime substitutes, when a continuous
spacetime manifold is substituted by a graph (Finkelstein, 1996) or
a finite topological space (Sorkin, 1991). It will be shown that any
finite-dimensional Hilbert space can be considered superspace if it
is endowed with an additional operation of associative product.

A 'quantum room' for finite topological spaces is built in this
paper. The sketch of the presented quantization scheme looks as
follows. At start we have a Hilbert space of states equipped with
the additional structure of associative algebra. Then an observable
is imposed splitting the space into its eigen-subspaces.  The core
of the scheme is the {\em spatialization procedure}: being applied
to a subspace, it manufactures a finite topological space, which is
interpreted as a spacetime substitute.

The main technical result is that if the state space of a
quantum system (with a finite number of degrees of freedom) is
endowed by an additional structure of algebra, then with any
observable property of the system we can associate a finite
topological space using the spatialization procedure. Non-trivial
results of this procedure occur when the algebra is noncommutative
(see the example in section \ref{sexample} below).  The account of
the results is organized as follows.

Section \ref{s1}: Finitary spacetime substitutes. Overviews the
discretization procedure (due to Sorkin, 1991) when continuous
manifolds are replaced by finite topological spaces.

Section \ref{s2}: Incidence algebras. Shows how to associate a
noncommutative algebra with any finite topological space.

Section \ref{s3}: The spatialization procedure. The inverse
operation is realized: having a finite-dimensional algebra on its
input, it furnishes a finite topological (or partially ordered)
set. Being applied to the incidence algebra of a poset, it restores
the initial partial order up to an order isomorphism.

Section \ref{s4}: Finitary algebraic superspace. Introduces the
super-(state space), or, in other words, the 'quantum room' for
finitary spacetime substitutes.

Section \ref{snoncomm}: Liaisons with non-commutative geometry.
Some ideas are presented to show where the dynamical equations for
the evolution of the spacetime topology could be taken from.

Section \ref{sexample}. An example. A toy model is built
based on the algebra of $4\times 4$ matrices. It is shown how
different eigenstates of {\em one} observable are associated with
the spaces of different topological structure.

\section{Finitary spacetime substitutes}\label{s1}

The coarse-graining procedure described in this section
substitutes a continuous manifold by a directed graph. This
procedure was introduced and described in detail by Sorkin
(1991): here an alternative account of this scheme based on the
notion of convergence is presented.

\paragraph{Formal point of view.} When we are speaking of
spacetime as a manifold $M$, its mere definition assumes that we
have a covering $\atlas$ of $M$ by open subsets. The idea of
coarse-graining is to replace the existing topology of the manifold
$M$ by that generated by the covering $\atlas$. As a result, the
spacetime manifold acquires the cellular structure with respect to
$\atlas$, so that the events belonging to one cell are thought of
as operationally indistinguishable. Then, instead of considering
the set $M$ of all events we can focus on its finite subset
$X\subseteq M$ such that each cell contains at least one point of
$X$.

To be more precise, consider the equivalence relation $\equiv$ on
the points of $M$: $x\equiv y$ if and only if they belong to one
cell, or, more strictly:

\[
x\equiv y \quad\hbox{ if and only if }\quad \forall {\cal O} \in \atlas
\quad x\in
{\cal O} \Leftrightarrow y\in {\cal O}
\]

Taking the quotient $M/ \equiv$ we obtain a $T_0$-space which
is
called {\sc finitary substitute} of $M$ with respect to the
covering $\atlas$. The topology on $M/ \equiv$ is induced by
the
canonical projection $M\to M/ \equiv$. Consider some
examples.

\begin{figure}[]
\begin{picture}(120,30)
\thicklines
\llll{0}{0}{1}{0}{30}
\llll{0}{0}{0}{1}{30}
\llll{0}{30}{1}{0}{30}
\llll{30}{0}{0}{1}{30}

\thinlines

\llll{10}{0}{0}{1}{30}
\llll{0}{10}{1}{0}{30}

\mmmm{60}{0}{0}{10}{4}{1}{0}{10}
\mmmm{60}{0}{10}{0}{2}{0}{1}{10}
\mmmm{60}{20}{10}{0}{2}{0}{1}{10}

\mmmm{0}{0}{2.5}{0}{9}{1}{1}{10}
\mmmm{0}{10}{0}{2.5}{9}{1}{-1}{10}

\mmmm{0}{2.5}{60}{0}{2}{1}{1}{7.5}
\mmmm{0}{5}{60}{0}{2}{1}{1}{5}
\mmmm{0}{7.5}{60}{0}{2}{1}{1}{2.5}

\mmmm{2.5}{0}{60}{20}{2}{-1}{1}{2.5}
\mmmm{5}{0}{60}{20}{2}{-1}{1}{5}
\mmmm{7.5}{0}{60}{20}{2}{-1}{1}{7.5}

\mmmm{7.5}{30}{60}{0}{2}{1}{-1}{2.5}
\mmmm{5}{30}{60}{0}{2}{1}{-1}{5}
\mmmm{2.5}{30}{60}{0}{2}{1}{-1}{7.5}

\mmmm{22.5}{0}{40}{0}{2}{1}{1}{7.5}
\mmmm{25}{0}{40}{0}{2}{1}{1}{5}
\mmmm{27.5}{0}{40}{0}{2}{1}{1}{2.5}

\llll{60}{0}{1}{1}{10}
\llll{60}{30}{1}{-1}{10}

\put(76,25){\mbox{is ${\cal O}_x$}}
\put(76,5){\mbox{is ${\cal O}_y$}}
\put(-3,-3.2){\mbox{\small 0}}
\put(9,-3.2){\mbox{\small 1/3}}
\put(31,-3.2){\mbox{\small 1}}
\put(-3.2,29){\mbox{\small 1}}
\put(-6,9){\mbox{\small 1/3}}

\end{picture}
\caption{The covering of a piece of plane.}
\label{fcovplane}
\end{figure}
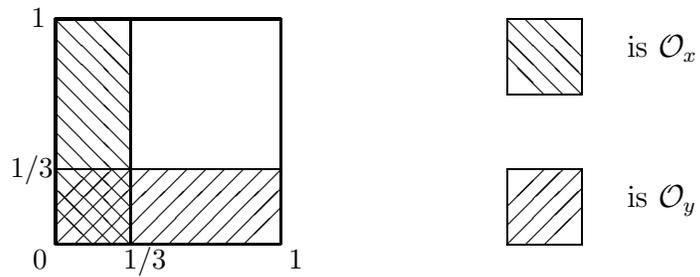

\paragraph{Example 1.} Let $M$ be a piece of plane:  $M=(0,1)\times
(0,1)$, and $\atlas = \{M, {\cal O}_x, {\cal O}_y\}$ be its
covering with ${\cal O}_x=(0,1/3)\times (0,1)$ and ${\cal
O}_y=(0,1)\times(0,1/3)$ (Fig. \ref{fcovplane})

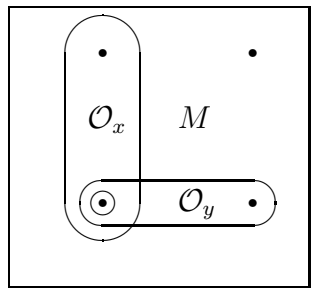
\begin{figure}[]
\begin{center}
\fbox{
\begin{picture}(35,35)
\multiput(10,10)(20,0){2}{\vertex}
\multiput(10,30)(20,0){2}{\vertex}

\put(10,10){\circle{3.2}}

\put(10,13){\line(1,0){20}}
\put(10,7){\line(1,0){20}}

\put(10,10){\oval(10,10)[b]}
\put(10,30){\oval(10,10)[t]}

\put(5,10){\line(0,1){20}}
\put(15,10){\line(0,1){20}}

\put(10,10){\oval(6,6)[l]}
\put(30,10){\oval(6,6)[r]}

\put(20,20){\mbox{$M$}}
\put(20,9){\mbox{${\cal O}_y$}}
\put(8,20){\mbox{${\cal O}_x$}}

\end{picture}
}
\end{center}
\caption{A finitary substitute of the plane.}
\label{fpls}
\end{figure}

The appropriate finitary substitute is presented in
Fig.\ref{fpls}.

\paragraph{ Example 2.} A circle (Balachandran {\em et al.,\/}
1996). Let $M=\exp(i\phi)$, and let the covering be $\atlas =
\{{\cal O}_1 ,{\cal O}_2, {\cal O}_3, {\cal O}_4\}$ with

\[
\begin{array}{lcl}
{\cal O}_1 = (-\pi/2,\pi) &;& {\cal O}_2 = (\pi/2,2\pi) \cr
{\cal O}_3 = (\pi/2,\pi) &;& {\cal O}_4 = (-\pi/2,0)
\end{array}
\]

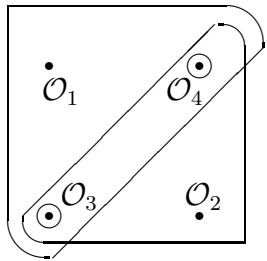
\begin{figure}[]
\begin{center}
\begin{picture}(40,40)(5,5)

\multiput(10,10)(20,0){2}{\vertex}
\multiput(10,30)(20,0){2}{\vertex}

\put(9,25.5){\mbox{${\cal O}_1$}}
\put(25.5,25.5){\mbox{${\cal O}_4$}}
\put(28,11.5){\mbox{${\cal O}_2$}}
\put(11.5,11.5){\mbox{${\cal O}_3$}}

\put(10,10){\oval(11,11)[lb]}
\put(10,10){\oval(7,7)[lb]}
\multiput(10,10)(20,20){2}{{\circle{3.2}}}

\put(6.5,10){\line(1,1){25.5}}
\put(10,6.5){\line(1,0){26}}
\put(36,6.5){\line(0,1){26}}

\put(10.5,4.5){\line(1,1){28}}
\put(4.5,10){\line(0,1){28}}
\put(32.5,32){\oval(7,7)[rt]}

\put(33,32.5){\oval(11,11)[rt]}
\put(4.5,38){\line(1,0){28.5}}

\end{picture}
\end{center}
\caption{A finitary substitute of the circle.}
\label{fcircle}
\end{figure}

The appropriate finitary substitute is shown in Fig.\ref{fcircle}.

\medskip

\addtocounter{footnote}{-2}
\renewcommand{\thefootnote}{\arabic{footnote}}

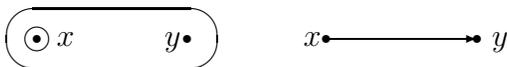
\begin{figure}[]
\[
\begin{array}{cc}
\begin{picture}(35,10)(3,3)
\put(10,10){\vertex}
\put(12.7,9){\mbox{$x$}}
\put(30,10){\vertex}
\put(27,9){\mbox{$y$}}

\put(10,10){{\circle{3.2}}}
\put(10,10){{\oval(8,8)[l]}}
\put(30,10){{\oval(8,8)[r]}}

\put(10,14){\line(1,0){20.5}}
\put(10,6){\line(1,0){20.5}}

\end{picture}
&
\begin{picture}(33,10)(3,3)
\put(10,10){\vertex}
\put(7,9){\mbox{$x$}}
\put(30,10){\vertex}
\put(32,9){\mbox{$y$}}
\put(10,10){\vector(1,0){20}}
\end{picture}
\end{array}
\]
\caption{The set of two points with non-trivial topology and its
graph of convergence.}

\label{f2pts}
\end{figure}

\noindent {\bf The graphs of finitary substitutes\footnote{Another
equivalent version of the transition between graphs and finite
topological spaces was presented in (Sorkin, 1991; Zapatrin,
1993).}.} Let us consider the behavior of sequences of elements in
finite topological spaces. First it is worthy to mention that if a
finite topological space is Hausdorf, then its topology is
necessarily discrete ({\em i.e.\/} in a sense degenerate). Since we
are going to deal with non-trivial topologies, we should not expect
them to be Hausdorf. As a consequence, the theorem of the
uniqueness of the limit of a sequence will not be valid anymore.
Consider the simplest example. Let $X=\{x,y\}$ be a set of two
points with the topology depicted in Fig.\ref{f2pts}.

Now consider the sequence $x,x,\ldots \, x,\ldots$ evidently
having $x$ itself as a limit point. However, by the mere
definition of the limit, $y$ is a limit point of this sequence
as well! So, we also have $x,x,\ldots \, x,\ldots\to y$  , denote
it briefly $x\to y$.

\begin{figure}[]
\[
\begin{array}{cc}
\begin{picture}(20,20)
\put(10,0){{\vertex}}
\put(0,10){{\vertex}}
\put(20,10){{\vertex}}
\put(10,20){{\vertex}}
\put(9.5,0.5){\vector(-1,1){9}}
\put(10,0.5){\vector(0,1){19}}
\put(10.5,0.5){\vector(1,1){9}}
\put(0.5,10.5){\vector(1,1){9}}
\put(19.5,10.5){\vector(-1,1){9}}
\end{picture}
\hskip7em
&
\hskip7em
\begin{picture}(20,20)
\put(10,0){{\vertex}}
\put(0,10){{\vertex}}
\put(20,10){{\vertex}}
\put(10,20){{\vertex}}
\put(9.5,0.5){\vector(-1,1){9}}
\put(10.5,0.5){\vector(1,1){9}}
\put(9.5,19.5){\vector(-1,-1){9}}
\put(10.5,19.5){\vector(1,-1){9}}
\end{picture}
\cr
\hbox{a).}
&
\hbox{b).}
\end{array}
\]
\caption{The graphs of convergencies for the finitary substitutes
of a). a piece of plane and b). a circle }
\label{fconvgraphs}
\end{figure}
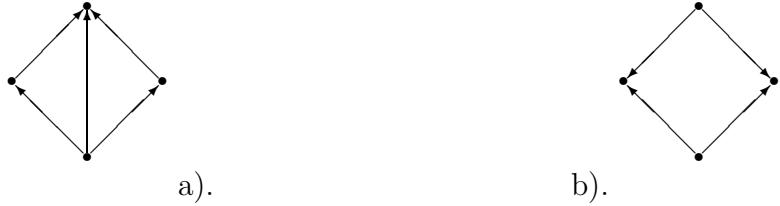

Another equivalent way to define a topology $\tau$ on a set $X$ is
to define which sequences in $X$ converge with respect to $\tau$
and which do not. So, for finite $X$ we can instead of drawing
pictures of open sets draw the graph of convergencies of sequences
$x,x,\ldots \, x,\ldots\to y$. In Fig. \ref{fconvgraphs} the
convergency graphs of the above examples are shown.

It is straightforward to prove that for any points $x,y,z,$ of a
finite topological space $X$ $x\to y$ and $y\to z$ imply $x\to
z$, therefore its graph of convergencies $G(X)$ will be always
{\em transitive}. Note that the transitivity holds only in the case
when the convergence is generated by a topology. For a more
detailed account of this issue the reader is referred to Isham
(1989).

\paragraph{From graphs to topological spaces.} Conversely, when we
have an arbitrary graph $G$ with the set of vertices $X$, we can
always define a topology $\tau(G)$ on $X$ as follows. The prebase
of $\tau(G)$ is the collection of all neighborhoods of the points
of $X$. A {\sc neighborhood} ${\cal O}(x)$ of $x\in X$ is defined
as the collection of all points of $y\in X$ from which $x$ is
reachable along the darts of the graph $G$, that is:

\[
{\cal O}(x)\,=\,\{y\in X\mid\: \exists y_0,\ldots,
y_n\in X\,:\, y_i\to y_{i+1}\, , \, y_0=y, \, y_n=x\}
\]

\begin{figure}[]
\begin{center}
\begin{picture}(67,25)(0,-5)
\put(0,0){\vertex}
\put(0,0){\vector(1,0){14.5}}
\put(15,0){\vertex}
\put(15,0){\vector(0,1){14.5}}
\put(15,15){\vertex}

\put(35,0){\vertex}
\put(35,0){\circle{3.2}}
\put(35,0){\oval(8,8)[l]}
\put(50,0){\oval(8,8)[r]}

\put(35,4){\line(1,0){15}}
\put(35,-4){\line(1,0){15}}
\put(50,0){\vertex}
\multiput(29,-7)(0,26){2}{\line(1,0){26}}
\multiput(29,-7)(26,0){2}{\line(0,1){26}}
\put(50,15){\vertex}

\put(70,0){\vertex}
\put(70,0){\vector(1,0){14.5}}
\put(70,0){\vector(1,1){14.5}}
\put(85,0){\vertex}
\put(85,0){\vector(0,1){14.5}}
\put(85,15){\vertex}
\end{picture}
\end{center}
\caption{The transition $G \to \tau(G) \to G(\tau(G))$}
\label{fnontrans}
\end{figure}
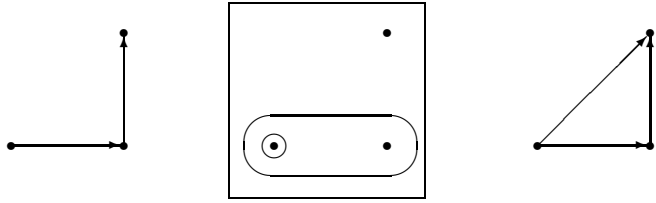

If the graph $G$ is transitive and we define the topology $\tau$ in
the way described above, the convergence graph of $\tau(G)$ will be
$G$ itself. In general, the convergency graph of the topology
$\tau(G)$ is the {\em transitive closure} of the graph $G$.
Consider one more example. Suppose we have a graph $G$ (Fig.
\ref{fnontrans}) which is not transitive. When we consecutively
pass from $G$ to $\tau(G)$ and then to $G(\tau(G))$ we obtain its
transitive closure (Fig. \ref{fnontrans})

\section{Incidence algebras}\label{s2}

\paragraph{Reminder on quasiorders and partial orders.} As it was
shown above, any finitary substitute can be associated with a
reflexive and transitive directed graph. When such a graph is set
up, we may consider its darts specifying a relation between
the points of $X$, denote it also $\to$. This relation has the
properties:

\[ \forall x\in X\qquad  x\to x \]

\begin{equation}
\forall x,y\in X \qquad (x\to y\hbox{ and }y\to z)\,
\hbox{ imply }\, x\to z
\label{qo}
\end{equation}

A relation on an arbitrary set having the properties (\ref{qo}) is
called {\sc quasiorder}. When a quasiorder is antisymmetric:

\begin{equation}
\forall x,y\in X\quad x\to y\hbox{ and }y\to x\hbox{  imply
}x=z
\label{po}
\end{equation}

\noindent the relation $\to$ is called {\sc partial order}. The
appropriate set will be called partially ordered or, for brevity,
{\sc poset}.

\paragraph{Incidence algebras.} I shall give here two equivalent
definitions of the notion of incidence algebra (Rota, 1968).
The first one will deal with posets in terms of
graphs, and the other one addresses directly to partial orders.

So, let $(X,\to)$ be a quasiordered set. Denote by the graph $G$ of
$(X,\to)$. Then consider the linear space $\algebra$ whose basis
${\bf e}_{ij}$ is labelled by the darts $(ij)$ of $G$, {\em i.e.\/}
by comparable pairs $i\to j$ in $X$.  Define the product in
$\algebra$ by setting it on its basic elements:

\begin{equation}\label{def22}
{\bf e}_{ij}{\bf e}_{jk}\quad =\quad
\left\lbrace\begin{array}{lcl}
{\bf e}_{il} &,& \hbox{if} \quad j=k \cr
0 &,& \hbox{otherwise}
\end{array}\right.
\end{equation}

Note that ${\bf e}_{il}$ in (\ref{def22}) is always well-defined
since $G$ is not an arbitrary graph but that of a partial order,
that is why the existence of darts $i\to j$ and $j\to k$ always
enables the existence of $i\to l$. The space $\algebra$ with the
product (\ref{def22}) is called the {\sc incidence algebra} of
the poset $(X,\to)$.

Another equivalent definition of the incidence algebra is the
following (Aigner, 1976). For a quasiordered set $X$ define its
incidence algebra $\algebra_X$, or simply $\algebra$ if no
ambiguity occurs, as the collection of all complex-valued functions
of two arguments vanishing on noncomparable pairs:

\begin{equation}\label{a}
\algebra=\{a:X\times X\to \cfield\mid a(x,y)\neq 0\Rightarrow
x\to
y\}\end{equation}

To make the defined linear space $\algebra$ algebra we define
the
product of two elements $a,b\in \algebra$ as:

\begin{equation}\label{aprod}
ab(x,y)=\sum_{z:x\to z\to y}^{} a(x,z)b(z,y)
\end{equation}

\noindent It can be proved that the defined product operation is
associative (Rota, 1968). Since the set $X$ is finite, the algebra
$\algebra$ is finite-dimensional associative (but not commutative,
in general) algebra over $\cfield$.

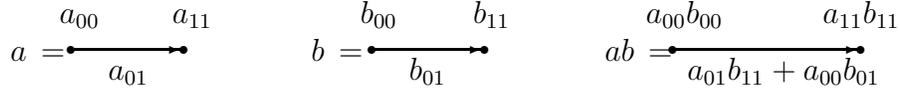
\begin{figure}[]
\begin{center}

\begin{picture}(108,25)(0,-5)
\put(-8,-1.5){\mbox{\(a\, =\, \)}}
\put(0,0){\vertex}
\put(-1.5,3.5){\mbox{\(a_{00}\)}}
\put(0,0){\vector(1,0){14.5}}
\put(5,-4){\mbox{\(a_{01}\)}}
\put(15,0){\vertex}
\put(13.5,3.5){\mbox{\(a_{11}\)}}

\put(32,-1.5){\mbox{\(b\, =\, \)}}
\put(40,0){\vertex}
\put(38.5,3.5){\mbox{\(b_{00}\)}}
\put(40,0){\vector(1,0){14.5}}
\put(45,-4){\mbox{\(b_{01}\)}}
\put(55,0){\vertex}
\put(53.5,3.5){\mbox{\(b_{11}\)}}

\put(71,-1.5){\mbox{\(ab\, =\, \)}}
\put(80,0){\vertex}
\put(76.5,3.5){\mbox{\(a_{00}b_{00}\)}}
\put(80,0){\vector(1,0){24.5}}
\put(82,-4){\mbox{\(a_{01}b_{11}+a_{00}b_{01}\)}}
\put(105,0){\vertex}
\put(100,3.5){\mbox{\(a_{11}b_{11}\)}}

\end{picture}
\end{center}
\caption{Transitions on a finitary substitute.}
\label{trans}
\end{figure}

Now let us clarify the meaning of the elements of $\algebra$.  Let
$a\in \algebra$ and $x,y$ be two points of $X$. If they are not
linked by a dart then, according to (\ref{a}), the value $a(x,y)$
always vanishes. So, $a(x,y)$ can be thought of as an assignment of
weights (or, in other words, transition amplitudes) to the darts of
the graph $X$. In these terms the product (\ref{aprod}) has the
following interpretation. Let $c=ab$, then $c(x,y)$ is the sum of
the amplitudes of all allowed two-step transitions, the first step
being ruled by $a$ and the second by $b$. As an example, consider
the set of two points (Fig. \ref{f2pts}). The result of
multiplication is shown on Fig. \ref{trans}.

åhe element $c(x,y)$ of the multiple product $c=a_1\ldots a_n$
looks similar to the Feynman sum over all paths from $x$ to $y$
of the length $n$ allowed by the graph $X$, making them similar
to $S$-matrices.

So, the transition from finitary substitutes to algebras is
described.
The inverse procedure of "spatialization" will be described below
in the Section \ref{s3}.

\paragraph{The standard matrix representation of incidence
algebras.} Given the incidence algebra of a quasiordered set $X$,
its standard matrix representation is obtained by choosing the
basis of $\algebra$ consisting of the elements of the form ${\bf
e}_{ab}$, with $ab$ ranging over all ordered pairs $a\to b$ of
elements of $X$, defined as:

\begin{equation}
{\bf e}_{ab}(x,y)= \cases{
1 & $x=a$ and $y=b$ (provided $a\to b$)\cr
0 & otherwise}
\label{eab}
\end{equation}

\noindent (such matrices are called matrix units). We can also
extend the ranging to {\em all} pairs of elements of $X$ by putting
${\bf e}_{ab}\equiv 0$ for $a\not\to b$\footnote{This is not strict
from the algebraic point of view since the collection of such ${\bf
e}_{ab}$ is not an ideal in the full matrix algebra.  However the
forthcoming results are valid for any finite topological space}.
Then the product (\ref{aprod}) reads:

\begin{equation}\label{mprod}
{\bf e}_{ab}{\bf e}_{cd} = \delta_{bc}{\bf e}_{ad}
\end{equation}

\noindent With each $a\in \algebra$ the following $N\times
N$-matrix ($N$ being the cardinality of the poset $X$) is
associated:

\[
a\mapsto a_{ik} = a(x_i,x_k)
\]

\noindent Let $I$ be the incidence matrix of the graph $X$, that
is

\begin{equation}\label{inc}
I_{ik} = \cases{
1 & $x_i\to x_k$\cr
0 & otherwise}
\end{equation}

\noindent then the elements of $\algebra$ are represented as the
matrices having the following property:

\begin{equation}\label{ai}
\forall i,k \quad a_{ik}I_{ik} = a_{ik}
\qquad\hbox{(no sum over }i,k )
\end{equation}

\noindent The product $c=ab$ of two elements is the usual
matrix product:

\[
c_{ik} = c(x_i,x_k) = \sum_{i\to l\to k}^{}a(x_i,x_l)b(x_l,x_k)
=
\sum_{\forall l}^{} a_{il}b_{lk}
\]

That means, we have so embedded $\algebra$ into the full matrix
algebra ${\cal M}_N(\cfield)$, that $\algebra$ is represented by
the set of all matrices satisfying (\ref{ai}). So, to specify an
incidence algebra in the standard representation we have to fix the
template matrix replacing the unit entries in $I_{ik}$ (\ref{inc})
by wildcards $*$ ranging independently over all numbers. We can
always re-enumerate the elements of $X$ to make $I_{ik}$ (and
hence the template matrix) {\em upper-block-triangular} matrix with
the blocks corresponding to cliques. In particular, when $X$ is
partially ordered, each clique contains exactly one element of $X$,
and the incidence matrix $I$ is upper triangular.

\paragraph{Examples.} Let us again return to our examples and
build the incidence algebras associated with the finitary
substitutes for the piece of plane and the circle. The template
matrices for the standard matrix representation of the
appropriate incidence algebras will have the form:

\begin{equation}\label{exofalg}
I_{\hbox{plane}} = \left(\begin{array}{cccc}
\ast & \ast & \ast & \ast \cr
0 & \ast & 0 & \ast \cr
0 & 0 & \ast & \ast \cr
0 & 0 & 0 & \ast \cr
\end{array}\right)
\quad ;\quad
I_{\hbox{circle}} = \left(\begin{array}{cccc}
\ast & 0 & 0 & 0 \cr
0 & \ast & 0 & 0 \cr
\ast & \ast & \ast & 0 \cr
\ast & \ast & 0 & \ast \cr
\end{array}\right)
\end{equation}

\medskip

\noindent where the wildcard $*$ denotes the ranging over the field
of numbers, for instance

\[
\left( \begin{array}{cc}
* & * \\
0 & *
\end{array} \right)
\,=\,
\left\lbrace \left.
\left( \begin{array}{cc}
a & b \\
0 & c
\end{array} \right)
\right\vert
\;
a,b,c \in \cfield
\right\rbrace
\]

\section{The spatialization procedure}\label{s3}

This section introduces the procedure which is inverse to that
described in section \ref{s2}. Namely, having a finite-dimensional
algebra on its input, the suggested spatialization procedure
manufactures a quasiordered set. Being applied to the incidence
algebra ${\cal A}_X$ of a quasiordered space $X$, it yields the
initial space $X$ (up to an isomorphism).

\paragraph{Imploding/exploding in quasiorders.} Let $(Y,\to )$ be a
quasiordered set (\ref{qo}). Define the relation $\sim$ on $Y$

\[
x\sim y\qquad\Leftrightarrow\qquad x\to y\hbox{ and }y\to x
\]

\noindent being equivalence on $Y$, and consider the quotient set
$X=Y/\sim$. Then $X$ is the partially ordered set (Birkhoff, 1967).
We shall call this procedure {\sc imploding} of a quasiorder.

When $Y$ is treated as the graph of a finitary substitute, the
transition from $Y$ to $X$ has the following meaning: $X$ is
obtained from $Y$ by smashing cliques to points. Contemplating this
procedure we see that $X$ may also be treated as the subgraph
obtained from $Y$ by deleting (except one from every
clique) 'redundant' vertices with adjacent (both incoming and
outgoing) darts.

We shall also consider the inverse procedure of {\sc exploding} a
partially set $X$ to a quasiorder $Y$. To each point of $x\in X$ a
positive integer $n_x$ is assigned. This number can be thought of
as inner dimension of infraspace (Finkelstein, 1996) ---
a room for gauge transformations.  Then each $x$ is replaced by its
$n_x$ copies linked between each other by two-sided darts and
linked with other vertices in the same way as $x$.

\begin{figure}[]
\begin{center}
\begin{picture}(80,20)

\put(-11,7){\mbox{\(Y\,=\,\)}}

\multiput(0,0)(0,15){2}{\mbox{\vertex}}
\multiput(15,0)(0,15){2}{\mbox{\vertex}}

\multiput(7.5,0)(0,15){2}{\vector(1,0){7}}
\multiput(7.5,0)(0,15){2}{\vector(-1,0){7}}
\multiput(0,0.5)(15,0){2}{\vector(0,1){14}}

\put(30,10){\vector(1,0){25}}
\put(35,11.5){\mbox{imploding}}

\put(55,5){\vector(-1,0){25}}
\put(35,1.5){\mbox{exploding}}

\put(69,7){\mbox{\(X\,=\,\)}}

\multiput(80,0)(0,15){2}{\mbox{\vertex}}
\put(80,0.5){\vector(0,1){14}}

\end{picture}
\end{center}
\caption{Imploding of a quasiorder vs. exploding a partial order.}
\label{fimplexpl}
\end{figure}
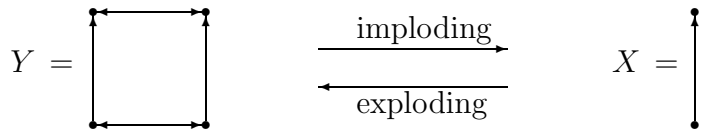

So, given a quasiordered set $Y$, we can always represent it as the
partially ordered set $x$ of its cliques equipped with the
additional structure: to each $x\in X$ an integer $n_x\ge 1$
thought of as the cardinality of appropriate clique is assigned:

\begin{equation}\label{qopo}
Y = (X, \{n_x\})
\end{equation}

\noindent which is illustrated in Fig.\ref{fimplexpl}.

\paragraph{Unformal reminder on Gel'fand techniques.} Suppose we
have a commutative $*$-algebra $\algebra$ and want to represent
it in the most 'natural' way, that is, by functions on a
topological space $M$.  Let us look at the algebra of continuous
functions $\algebra = C(M)$. Select a point $m\in M$, and consider
the collection

\[
k_m = \{f\in \algebra \mid f(m)=0\}
\]

We see that $k_m$ is an ideal in $\algebra$, and the factor algebra
$\algebra / k_m$ is one dimensional, hence it is simple. The term
{\em simple} has its strict mathematical definition: an algebra $A$
is called simple if it has no proper ({\em i.e.\/} different from 0
and $A$) ideals. An ideal $\ideal$ of an algebra $\algebra$ is
called primitive if the factor algebra $\algebra/\ideal$ is simple.
In the case when $M$ is a 'good' space, there is 1--1
correspondence between the points of $M$ and primitive ideals. So,
for commutative algebras the Gel'fand spatialization procedure
looks as follows:

\[
\spat(\algebra) = \{\,\hbox{primitive ideals}\,\}
\]

\noindent Now, following Landi (1997), let us pass to the
noncommutative case.  For technical reasons I will use another
equivalent definition of primitive ideal. An ideal $x$ of an
algebra $\algebra$ is called {\sc primitive} if it is the kernel of
an irreducible representation of $\algebra$ in a vector space
$V_x$.  For finite dimensional algebras the dimension of $V_x$ does
not depend on a particular representation of $\algebra$.

\paragraph{The spatialization procedure.} A construction which
builds quasiordered sets by given finite-dimensional algebras is
described here. Let $\algebra$ be a subalgebra of the full matrix
algebra ${\rm Mat}_n(\cfield)$.  To build the quasiordered set
associated with $\algebra$ the following is to be performed.

\bigskip

\noindent {\sc Step 1. Creating cliques.} Just as in the Gel'fand
theory consider the set of all primitive ideals of $\algebra$. For
every such ideal $x$ denote by $n_x$ the dimension of the
appropriate representation space $V_x$

\[ n_x = \dim V_x \]

\noindent Then declare the set

\[
X\,=\, {\rm Prim}\,\algebra \,=\,\{ x \mid
\,x\, \hbox{is a primitive ideal in}\, \algebra \}
\]

\noindent to be the set of cliques, and the numbers $n_x$ to be
their cardinalities. So, the future finitary substitute is already
created as a set (or, more precisely, as an equipped set since we
admit inner dimensions $n_x$ of its points). It remains to endow
$X$ with the appropriate topology.

\medskip

\noindent {\sc Step 2. Stretching the elementary darts.} For every
pair $x,y\in X$ we form their product $xy$:

\[
xy=\{a\in \algebra \mid
\, \exists u\in x,\, v\in y:\, uv=a\}
\]

\noindent and their set intersection $x\cap y$. Both $xy$ and
$x\cap y$ are ideals in $\algebra$ and

\begin{equation}\label{e2}
xy\subseteq x\cap y
\end{equation}

\noindent (since both $x$, $y$ are ideals), however the reverse
inclusion may not hold. The rule I suggest is the following: the
dart $x\to y$ is stretched if and only if the inclusion (\ref{e2})
is proper:

\begin{equation}\label{qxy}
x\to y\quad
\hbox{if and only if} \quad
xy\neq x\cap y
\end{equation}

\medskip

\noindent {\sc Step 3. Forming the partial order.} When (\ref{qxy})
is checked for all pairs $x,y$, the nontransitive predecessor of the
partial order on the set $X$ is obtained. To make $X$ partially
ordered form the transitive closure of the obtained relation:

\[
{\rm darts}(X) := \{(x,x)\}_{x\in X}\,\cup \,
\{(x,z)\mid \exists x =
y_0,
\ldots ,y_n = z\,\, Q(y_i,y_{i+1})\neq 0\}
\]

So, the finitary substitute $Y = (X,n_x)$ (\ref{qopo}) is
completely built. In the sequel denote the quasiordered set $Y$
furnished by the spatialization procedure applied to the algebra
$\algebra$ by

\begin{equation}\label{dspat}
Y = \spat\algebra
\end{equation}

\medskip

A remarkable property of the incidence algebras is the following.
Being applied to the incidence algebra of a quasiorder $Y$, this
procedure restores $Y$ up to an isomorphism of quasiorders. This
was proved by Stanley (1986) for partial orders, however his prove
survives for quasiorders as well. To see how it works, consider an
example.

\medskip

\paragraph{An example.} Let us explicitly restore the quasiorder
associated with the finitary substitute of the plane
(Figs.\ref{fcovplane},\ref{fpls}). So, the algebra $\algebra$ is
the collection of matrices of the following form (\ref{exofalg}):

\[
\algebra \,=\,
\left(\begin{array}{cccc}
\ast & \ast & \ast & \ast \cr
0 & \ast & 0 & \ast \cr
0 & 0 & \ast & \ast \cr
0 & 0 & 0 & \ast \cr
\end{array}\right)
\]

\noindent where, as above, the wildcard $\ast$ ranges over all
numbers.  Now let us perform the spatialization procedure step by
step.

\medskip

\noindent {\sc Step 1.} ${\cal K}$ has exactly 4 characters, denote
their kernels by 1,2,3,4:

\begin{equation}\label{e4ker}
\begin{array}{c}
1= \kernel\chi_1 \, = \,
\left(\begin{array}{cccc}
0 & \ast & \ast & \ast \cr
0 & \ast & 0 & \ast \cr
0 & 0 & \ast & \ast \cr
0 & 0 & 0 & \ast \cr
\end{array}\right)
\quad ; \quad
2= \kernel\chi_2 \, = \,
\left(\begin{array}{cccc}
\ast & \ast & \ast & \ast \cr
0 & 0 & 0 & \ast \cr
0 & 0 & \ast & \ast \cr
0 & 0 & 0 & \ast \cr
\end{array}\right)
\cr \cr
3= \kernel\chi_3 \, = \,
\left(\begin{array}{cccc}
\ast & \ast & \ast & \ast \cr
0 & \ast & 0 & \ast \cr
0 & 0 & 0 & \ast \cr
0 & 0 & 0 & \ast \cr
\end{array}\right)
\quad ; \quad
4= \kernel\chi_4 \, = \,
\left(\begin{array}{cccc}
\ast & \ast & \ast & \ast \cr
0 & \ast & 0 & \ast \cr
0 & 0 & \ast & \ast \cr
0 & 0 & 0 & 0 \cr
\end{array}\right)
\end{array}
\end{equation}

\noindent so the set of cliques is $X=\{1,2,3,4\}$. All the cliques
have dimension one since all the representations $\chi_i$ are
one-dimensional.

\medskip

\noindent {\sc Step 2.} Now let us see how (\ref{qxy}) works. To
perform the calculations, the ordinary matrix product is used with
the following arithmetics: $\ast\cdot \ast = \ast$; $0\cdot \ast =
0$.  Take two points, say 1 and 2, then

\[
1\cap 2 \,=\,
\left(\begin{array}{cccc}
0 & \ast & \ast & \ast \cr
0 & 0 & 0 & \ast \cr
0 & 0 & \ast & \ast \cr
0 & 0 & 0 & \ast \cr
\end{array}\right)
\quad ; \quad
1\cdot 2 \,=\,
\left(\begin{array}{cccc}
0 & 0 & \ast & \ast \cr
0 & 0 & 0 & \ast \cr
0 & 0 & \ast & \ast \cr
0 & 0 & 0 & \ast \cr
\end{array}\right)
\]

\noindent so, $1\cap 2 \neq 1\cdot 2$, while

\[
2\cdot 1 \,=\,
\left(\begin{array}{cccc}
0 & \ast & \ast & \ast \cr
0 & 0 & 0 & \ast \cr
0 & 0 & \ast & \ast \cr
0 & 0 & 0 & \ast \cr
\end{array}\right)
\,=\,
1\cap 2
\]
which means that we have the arrow $1\to 2$, but not the reverse
$2\to 1$.  All the rest of calculation is performed quite
analogously.
\smallskip

\noindent {\sc Step 3.} Finally we get the following elementary
darts:  $1\to 2$, $1\to 3$, $2\to 4$, $3\to 4$. To complete these
darts to a partial order, it remains to add only one dart $1\to 4$
and the loops $1\to 1,\ldots,4\to 4$.  So, the poset (Fig.
\ref{fconvgraphs}.a) reproducing the plane is restored.

\section{Finitary algebraic superspace}\label{s4}

\paragraph{Topokinematics.} Let ${\cal H}$ be the state space of a
physical system. If we admit that besides its usual Hilbert
structure the space ${\cal H}$ is equipped with a product operation
$(\cdot):{\cal H}\times {\cal H}\to {\cal H}$ then {\em any}
observable can be thought of as a kind of
"topologimeter"\footnote{The origin of the product in the state
space is investigated in (Parfionov and Zapatrine, 1997)}. Namely,
let ${\bf A}$ be an observable, denote by $A$ the self-adjoint
operator in ${\cal H}$ associated with it.  $A$ splits ${\cal H}$
into the sum of its eigen-subspaces:

\begin{equation}\label{e26}
{\cal H} = {\cal H}_1\oplus \ldots \oplus {\cal H}_n
\end{equation}

According to the projection postulate, when the measurement of
${\bf A}$ is performed, any state vector $\psi \in {\cal H}$
collapses into one of the subspaces ${\cal H}_i$ (\ref{e26}) with
the probability

\begin{equation}\label{e271}
p_i=\mid(P_i,\psi)\mid^2
\end{equation}

\noindent where $P_i$ is the projector onto ${\cal H}_i$. In the
meantime, if a particular ${\cal H}_i$ is chosen, we may span on
it a subalgebra $\algebra_i = \algspan {\cal H}_i$:

\begin{equation}\label{e272}
\algebra_i = \algspan {\cal H}_i =
\cap\{{\cal H}'\mid {\cal H}_i\subseteq {\cal H}'\, ;
\, {\cal H}' \hbox{ is a subalgebra of }{\cal H}\}
\end{equation}

That means that measuring the observable ${\bf A}$ we obtain
subalgebras $\algebra_i$ (\ref{e272}) with probabilities $p_i$
(\ref{e271})\footnote{It should be, however, remarked that for
different ${\cal H}_i$ their spans $\algebra_i$ may coincide or
be isomorphic, so we have to adjust the correspondence ${\cal
A}_i\to p_i$ by summing the probabilities of the isomorphic
algebras}. The next step is to interpret each $\algebra_i$ as
spacetime, but this is what the spatialization procedure does:

\begin{equation}\label{e273}
X_i = \spat (\algebra_i)
\end{equation}

\noindent is a finite topological space associated with the
subalgebra $\algebra_i$. So, the quantum detection of topology is
described by the following quantization scheme:

\[
\left(\begin{array}{rc}
&\hbox{observables}\cr
+& \cr
&\hbox{states}
\end{array}\right)
\,\to\, \hbox{eigenspaces}\,\to\,
\hbox{subalgebras}\,\to\, \hbox{topological spaces}
\]

\paragraph{Two approaches to topodynamics.} Within the suggested
framework, the dynamics of topology can be introduced in two
ways being similar to the Heisenberg and Schr\"odinger pictures
in quantum mechanics.

The first approach is to fix a state $\psi \in {\cal H}$ and vary
the constants of multiplication which make the state space ${\cal
H}$ algebra. In this case the dynamical equations of evolution of
the spacetime topology is referred to equations for these
constants (note that there is a confusion in terms: in this
context the co\-n\-s\-tants-of-mu\-l\-ti\-p\-li\-ca\-tion are not
constants at all being subject to variation). However, this
opportunity will not be considered in this paper.

Another approach is the following. We have to invent some
topodynamical equation for the state vector $\psi$ under the
assumption that the algebraic ({\em i.e.} product) structure of the
state space ${\cal H}$ is fixed. For the case when ${\cal H}$ is
the full matrix algebra there is a remarkable link with the
noncommutative geometry (Zapatrin, 1995) which is
considered below (Section \ref{snoncomm}).

\section{Liaisons with non-commutative geometry}\label{snoncomm}

In the standard general relativity the metrics on a {\em given
manifold} is subject to variation, and the equations which
control it are the Einstein equations:

\begin{equation}\label{e30}
R_{ik} - \frac{1}{2}Rg_{ik} = -GT_{ik}
\end{equation}

It was shown by Geroch (1972) that in order to write down
(\ref{e30}) we do not need the manifold as a set, since the
entire contents of general relativity can be reformulated in
terms of the algebra $\algebra$ of smooth functions on the
underlying manifold. The vectors and the tensors are then
introduced in mere terms of $\algebra$ treated as algebra
(Parfionov and Zapatrin, 1995).  In this situation the algebra
$\algebra$ is called Einstein algebra.  Being algebra of functions
on a set, it is commutative.

Then we may pass to noncommutative Einstein algebras (Heller,
1995). We can still write down the equation (\ref{e30}), and
even try to solve it for some special cases (Zapatrin, 1995),
but the problem of its physical meaning immediately arises.

Now let us again return to the commutative case. Suppose the
equation (\ref{e30}) is solved and, as a result, we have the metric
tensor $g_{ik}$. Having this tensor we can, in turn, at each point
$m\in M$ consider it as a quadratic form and split out its $n$
eigenvectors. Having these eigenvectors we can then restore a part
of the manifold $M$ as exponential neighborhood of the point $m$.
This is in a sense classical spatialization.  What could be an
analog of such a consecutive construction in the noncommutative
case? The problem is that the equation (\ref{e30}) entangles {\em
vectors and tensors} which are not the elements of the algebra
$\algebra$ itself. However, when $\algebra$ is the incidence
algebra of a poset, this problem can be solved.

\paragraph{Generalities.} The idea of noncommutative geometry is to
replace the (commutative) algebra of smooth functions on a manifold
by a noncommutative algebra and then to build an algebraic analog
of usual differential geometry (Dubois-Violette, 1981). This is
possible due to the fact that the principle objects of
differential geometry can be reformulated in mere terms of algebras
of smooth functions. To see it, let us dwell on the vector
calculus.

In standard differential geometry the vector fields on a manifold
$M$ are in 1--1 correspondence the derivations of the algebra
$\algebra$ of smooth functions on $M$. Recall that a derivation in
an algebra $\algebra$ is a linear mapping $\vctr:  \algebra \to
\algebra$ enjoying the Leibniz rule:

\[
\vctr (ab)= \vctr a\cdot b+a\cdot \vctr b
\]

\noindent for all $a,b\in \algebra$. Denote by $\der \algebra$ the
set of all derivations in $\algebra$. In classical geometry the
correspondence between a vector field and the appropriate
derivation looks as follows:

\[
\vctr f = \sum_i v^i \frac{\partial f}{\partial x^i}
\]

So, the geometrical notion of vector field can be equivalently
replaced by the purely algebraic notion of derivation. The starting
point to build the differential geometry could be to choose an
appropriate algebra $\algebra$ called {\sc basic algebra}
(Parfionov and Zapatrin, 1995).

\paragraph{Noncommutative situation.} When the basic algebra
$\algebra$ is noncommutative there always exist a class of
derivations called {\sc inner} ones which are associated with the
elements $a\in \algebra$ in the following way: $a\mapsto \vec{a}
\in \der \algebra$ such that

\[
\vec{a}(b) = [a,b] = ab-ba
\]

The notorious property of any incidence algebra (and of full
matrix algebra in particular) is that any its derivative is inner.
That means that for any vector $v\in \der \algebra$ there exists
$\hat{v}\in \algebra$ such that for any $a\in \algebra$ the
vector $v$ acts as the commutator:

\begin{equation}\label{e32}
va = [\hat{v},a]
\end{equation}

In this case, if we obtain a solution of the Einstein equation
(\ref{e30}) associated with a subspace $V$ of vectors, we can
immediately bring this $V$ into the algebra $\algebra$:

\[
V\,\mapsto\, \hat{V}\,\subseteq\, \algebra
\]

\noindent and then directly apply the spatialization procedure
described in section \ref{s3} above. An example of how it can be
done is in the next section.

\section{An example}\label{sexample}

In this section I present an explicit example of 16-dimensional
Hilbert space endowed with a product operation, and a self-adjoint
operator in this space such that for two its different eigenspaces
the spatialization procedure yields two topologically different
spaces: a piece of plane and a circle.

Let ${\cal H}$ be the space of all complex valued $4\times 4$
matrices:

\[ {\cal H} = \mata_4 \]

The additional product operator on ${\cal H}$ will be the usual
matrix product. Define the following scalar product on ${\cal H}$:
for any $a,b \in {\cal H}$

\begin{equation}\label{e10}
<\!a,b\!> = \tr(agb^{\dag})
\end{equation}

\noindent where $\tr$ is the usual matrix trace, $(\,)^{\dag}$ is
the matrix transposition and
\[ g \,=\, \left(\begin{array}{cccc} 0
& 1 & 1 & 0 \cr 1 & 0 & 0 & 0 \cr 1 & 0 & 1 & 0 \cr 0 & 0 & 0 & 1
\end{array}\right) \]

The metric $<,\!>$ is nondegenerate, however indefinite: for any
$a\in {\cal H}$ the value

\[
<\!a,a\!> = r{\rm
Re}\left(\sum_i\overline{a_{i1}}(a_{i2}+a_{i3})\right) +
\sum_i(\vert a_{i3} \vert^2 + \vert a_{i4}\vert^2)
\]

\noindent is always real, but may have an arbitrary sign.

Consider the operator $N: {\cal H} \to {\cal H}$ defined as follows:

\begin{equation}\label{e1188}
Na = an
\end{equation}

\noindent with the following matrix $n$:

\[
n \,=\,
\left(\begin{array}{cccc}
2 & 0 & 0 & 0 \cr
0 & 2 & 0 & 0 \cr
0 & 1 & 1 & 0 \cr
0 & 0 & 0 & 1
\end{array}\right)
\]

The operator $N$ is self-adjoint. To verify it first note that
$ng = gn^{\dag}$, then

\[
<\!Na,b\!> = \tr(angb^{\dag}) =
\tr(agn^{\dag} b^{\dag}) = \tr(ag(bn)^{\dag}) =
<\!a,Nb\!>
\]

Now consider two vectors $p,c\in {\cal H}$:

\[
c \,=\,
\left(\begin{array}{cccc}
0 & 0 & 0 & 0 \cr
0 & 0 & 0 & 0 \cr
1 & 1 & 0 & 0 \cr
1 & 1 & 0 & 0
\end{array}\right)
\quad ; \quad
p \,=\,
\left(\begin{array}{cccc}
0 & 1 & 1 & 0 \cr
0 & 0 & 0 & 1 \cr
0 & 0 & 0 & 1 \cr
0 & 0 & 0 & 0
\end{array}\right)
\]

The direct calculations show that the vectors $p,c$ are
eigenvectors for the operator $N$ (\ref{e1188}) with the
eigenvalues 1 and 2, respectively:

\[
Nc = c
\quad ; \quad
Np = 2p
\]

Now recall that ${\cal H}$ has the structure of algebra, and consider
the subalgebras in ${\cal H}$ spanned on $p$ and $c$, respectively.
They are:

\[
\algspan\, p \,=\,
\left(\begin{array}{cccc}
\ast & \ast & \ast & \ast \cr
0 & \ast & 0 & \ast \cr
0 & 0 & \ast & \ast \cr
0 & 0 & 0 & \ast
\end{array}\right)
\quad ; \quad
\algspan\, c \,=\,
\left(\begin{array}{cccc}
\ast & 0 & 0 & 0 \cr
0 & \ast & 0 & 0 \cr
\ast & \ast & \ast & 0 \cr
\ast & \ast & 0 & \ast
\end{array}\right)
\]

\noindent having the dimensions 9 and 8, respectively. Then, if
we apply the spatialization procedure (section \ref{s3}), we
obtain two finite topological spaces

\begin{figure}[]
\[
\begin{array}{cc}
\begin{picture}(20,10)(0,10)
\put(10,0){{\vertex}}
\put(0,10){{\vertex}}
\put(20,10){{\vertex}}
\put(10,20){{\vertex}}
\put(9.5,0.5){\vector(-1,1){9}}
\put(10,0.5){\vector(0,1){19}}
\put(10.5,0.5){\vector(1,1){9}}
\put(0.5,10.5){\vector(1,1){9}}
\put(19.5,10.5){\vector(-1,1){9}}
\end{picture}
&
\begin{picture}(20,20)(0,10)
\put(10,0){{\vertex}}
\put(0,10){{\vertex}}
\put(20,10){{\vertex}}
\put(10,20){{\vertex}}
\put(9.5,0.5){\vector(-1,1){9}}
\put(10.5,0.5){\vector(1,1){9}}
\put(9.5,19.5){\vector(-1,-1){9}}
\put(10.5,19.5){\vector(1,-1){9}}
\end{picture}
\cr
\mbox{a).} & \mbox{b).}
\end{array}
\]
\caption{a). $\spat(\algspan\, p)$; b). $\spat(\algspan\, c)$}
\end{figure}

\noindent corresponding to a piece of plane (cf. Fig.
\ref{fconvgraphs} a.) and to the circle (cf.Fig. \ref{fconvgraphs}
b.)

\section*{Acknowledgments}

The major part of the ideas exposed in this paper was born and
elaborated in discussions with prof. G.Landi who organized my visit
to the University of Trieste (November--December, 1995) and enabled
its support from the Italian National Research Council. A partial
financial support from the Russian Foundation for Fundamental
Research (grant RFFI 96-02-19528) is appreciated.

\bigskip
\noindent{\bf\Large\bf References}
\medskip

Aigner, M., (1976),
{\it Higher Combinatorics,\/}
NATO ASI ser.C,
{\bf 31},
Berlin (West)
\smallskip

Balachandran A.P., G. Bimonte, E.Ercolessi, G. Landi, F. Lizzi,
G. Sparano, P. Teotonio-Sobrinho, (1996),
Noncommutative lattices as finite approximations,
{\it Journal of Geometry and Physics,\/}
{\bf 18}, 163--194,
(eprint hep-th/9510217)
\smallskip

Birkhoff, G, (1967),
{\it Lattice theory},
Providence, Rhode Island
\smallskip

Dubois-Violette, M., (1981),
Calcul diffe\'erentiel et g\'eometrie differentielle
non-commutative, {\it Comptes Rendus de l'Acad\'emie de Sciences de
Paris, ser.I, Math\'ematiques},
\smallskip

Finkelstein, D., (1996),
{\it Quantum Relativity,\/}
Springer, Berlin
\smallskip

Geroch, R., (1968),
What is a singularity in general relativity?
{\it Annals of Physics},
{\bf 48},
526
\smallskip

Geroch, R., (1972)
Einstein Algebras,
{\it Communications in Mathematical Physics,\/}
{\bf 26},
271
\smallskip

Heller, M., (1995),
Commutative and non-commutative Einstein algebras,
{\it Acta Cosmologica,\/}
{\bf XXI-2},
111
\smallskip

Isham, C.J., (1989),
An introduction to general topology and quantum topology.
In: Proceedings of the NATO ASI on Physics, Geometry and Topology
(August, 1989),
pp. 1--64
\smallskip

Isham, C.J., (1994),
Quantum logic and the histories approach to quantum
mechanics,
{\it Journal of Mathematical Physics },
{\bf 35},
2157
\smallskip

Landi, G., (1997),
An introduction to noncommutative spaces and their geometry,
181 pp., eprint hep-th/9701078
\smallskip

Misner, C., K.S.Thorne, and J.A.Wheeler, (1973),
{\it Gravitation,\/}
Freeman, San Francisco
\smallskip

Parfionov G.N., and R.R.Zapatrin, (1995),
Pointless Spaces in General Relativity,
{\it International Journal of Theoretical Physics},
{\bf 34},
717,
(eprint gr-qc/9503048)
\smallskip

Parfionov G.N., and R.R.Zapatrine, (1997),
Empirical topology in the histories approach to
quantum theory
(eprint gr-qc/9703011)
\smallskip

Pierce, R.P., (1982),
{\it Associative algebras,\/}
Springer, Berlin
\smallskip

Rota, G.-C., (1968),
On The Foundation Of Combinatorial Theory, I. The Theory Of
M\"obius Functions,{\it Zetschrift f\"ur
Wahrscheinlichkeitstheorie,\/}
{\bf 2},
340
\smallskip

Sorkin, R.D.,(1991),
Finitary substitutes for continuous topological spaces,
{\it International Journal of Theoretical Physics},
{\bf 30},
923
\smallskip

Stanley R.P., (1986),
{\it Enumerative combinatorics},
Wadsworth and Brooks, Monterrey, California
\smallskip

Zapatrin, R.R. (1993),
Pre-Regge Calculus: Topology Via Logic,
{\it International Journal of Theoretical Physics},
{\bf 32},
779
\smallskip

Zapatrin, R.R., (1995),
Matrix models for spacetime topodynamics,
{\it In: Procs of the ICOMM'95 (Vienna, June 3--6,
1995),\/}
1--19,
(eprint gr-qc/9503066)
\smallskip

\end{document}